\documentclass[reprint, superscriptaddress, secnumarabic, amssymb, nobibnotes, aps, prl]{revtex4-1}

\usepackage{graphicx}
\usepackage{epstopdf}
\usepackage[T1]{fontenc}
\usepackage[latin9]{inputenc}
\usepackage{amsbsy}
\usepackage{gensymb}
\usepackage[T1]{fontenc}
\usepackage[latin9]{inputenc}
\usepackage{amsmath}
\usepackage{amssymb}
\usepackage{bbm}
\usepackage{braket}
\usepackage{xcolor}
\allowdisplaybreaks
\usepackage{graphicx}
\usepackage[colorlinks=true]{hyperref}
\hypersetup{
    bookmarks=true,         
    unicode=false,          
    pdftoolbar=true,        
    pdfmenubar=true,        
    pdffitwindow=false,     
    pdfstartview={FitH},    
    pdftitle={Strongly anisotropic TaSeS system a new promising candidate for multi-gap superconductor},    
    pdfauthor={a},     
    pdfsubject={},   
    pdfcreator={},   
    pdfproducer={}, 
    pdfkeywords={} {} {}, 
    pdfnewwindow=true,      
    colorlinks=true,       
    linkcolor=blue, 
    citecolor=blue,        
    filecolor=magenta,      
    urlcolor=blue           
} 
\usepackage[normalem]{ulem}
\usepackage{amsmath}
\newcommand{\equref}[1]{Eq.~(\ref{#1})}

\newcommand{\figref}[1]{Fig.~\ref{#1}}

\newcommand{\tableref}[1]{Table~\ref{#1}}

\renewcommand{\approx}{\simeq}

\begin{document}
\title{\textrm{Two Dimensional Multigap Superconductivity in Bulk 2H-TaSSe}}

\author{C. Patra}
\affiliation{Department of Physics, Indian Institute of Science Education and Research Bhopal, Bhopal, 462066, India}
\author{T. Agarwal}
\affiliation{Department of Physics, Indian Institute of Science Education and Research Bhopal, Bhopal, 462066, India}
\author{Rajeshwari R. Chaudhari}
\affiliation{Department of Physics, Indian Institute of Science Education and Research Bhopal, Bhopal, 462066, India}
\author{R. P. Singh}
\email[]{rpsingh@iiserb.ac.in}
\affiliation{Department of Physics, Indian Institute of Science Education and Research Bhopal, Bhopal, 462066, India}

\begin{abstract}
\begin{flushleft}

\end{flushleft}
Superconducting transition metal dichalcogenides emerged as a prime candidate for topological superconductivity. This work presents a detailed investigation of superconducting and transport properties on 2H-TaSeS single crystals using magnetization, transport, and specific heat measurements. These measurements suggest multigap anisotropic superconductivity with the upper critical field, breaking Pauli limiting field in both in-plane and out-of-plane directions. The angle dependence of the upper critical field suggests 2-dimensional superconducting nature in bulk 2H- TaSeS.
\end{abstract}
\maketitle
\section{INTRODUCTION}
Recently superconductivity in layered transition metal dichalcogenides (TMDs) \cite{ CDW,CDW_in_NbSe2, structure,graphene_structure} have drawn immense research interest due to its ability to host unique physical properties such as superconductivity, charge density wave (CDW) \cite{fermisurface_nesting,polar_charge_symme_TaS2}, electron-electron correlation \cite{electronic_corelation}, and topological properties \cite{topology_case}. TMDs with a general formula MX$_2$ (M = Mo, W, Ta, Nb and X = Te, Se, S) consist of a layer of TM atom sandwiched between layers of chalcogen atoms and can exist in different polymorphs such as hexagonal 2H, trigonal 1T, and rhombohedral 3R structures \cite{graphene_structure,structure_differ}. Among the different families of TMDs, archetypal systems, NbSe$_2$ \cite{structure_NbSe2}, NbS$_2$ \cite{NbS2_structure}, TaSe$_2$ \cite{1T_TaSe2_xSx_pressure}, TaS$_2$ \cite{TaS2_structure} are reported to be intrinsic superconductors, while superconductivity has been successfully induced by the application of pressure \cite{pressure_SC_MoTe2,SC_doping, SC_doping1} and chemical doping in semiconducting MoS$_2$ \cite{MoS2}, MoSe$_2$ \cite{SC_MoSe2} and WTe$_2$ \cite{WTe2_SC}.\\

The Ta-based TMD system, 2H-TaSe$_2$ and 2H-TaS$_2$, is known to host charge density wave and superconductivity at low temperatures \cite{TS2,disorder_TaSeS}. Moreover, 2H-TaSe$_2$ shows CDW ordering at 90 K and the onset of superconductivity at 0.15 K \cite{CDW_TaSe2,TaSe2_xSx}, 2H-TaS$_2$ exhibits a chiral charge order system with superconductivity below 0.8 K \cite{TS2,TaS2_chairal}. The enhancement of the superconducting transition in monolayers and intercalation of elements or organic compounds \cite{mono_layer_TaS2, CuxTaS2, mono_TaSe2} resulted from the suppression of CDW by doping or disorder.

Recently 4H$_b$-TaS$_2$ phase consists of alternating stacking of weakly coupled 1T-TaS$_2$ (Mott insulator and proposed gapless spin liquid \cite{spin_liquid_TaS2}) and 1H-TaS$_2$ (2D superconductor with charge density wave) revealed the signature of time-reversal symmetry breaking (TRSB) \cite{TRSB_4H_TaS2} in the superconducting ground state. Apart from this, the different phase of TaSeS shows exiting superconducting properties. Scanning tunnelling microscopic studies on the disorder-driven superconductor 1T-TaSeS suggest a non-trivial link between superconductivity and charge order \cite{STM_TaSeS,topology_Tas2}. 2H-TaSeS shows an enhanced superconducting transition temperature, resulted from suppression of CDW due to disorder, as the similar electronic structure of 2H-TaSe$_2$, 2H-TaSeS and 2H-TaS$_2$ ruled out the role of the dopant \cite{disorder_TaSeS, electronic_sturcture_of_2H_TaS2}. However, the detailed superconducting, anisotropic and normal state properties of different phases of TaSeS are not available, which are crucial to understanding the superconducting gap/ground state of any superconductors \cite{disorder_TaSeS}\\

This work reports single crystal growth and detailed electronic and superconducting properties of 2H-TaSeS. Comprehensive critical field and heat capacity measurements suggest anisotropic multigap superconducting behaviour. The large in-plane and out-of-plane directions upper critical field breaks the Pauli limiting field. Interestingly, the angle dependence of the upper critical field well-fit 2D Tinkham model suggests 2D superconductivity in bulk 2H-TaSeS single crystal. 
\begin{figure*}
\includegraphics[width=2.0\columnwidth]{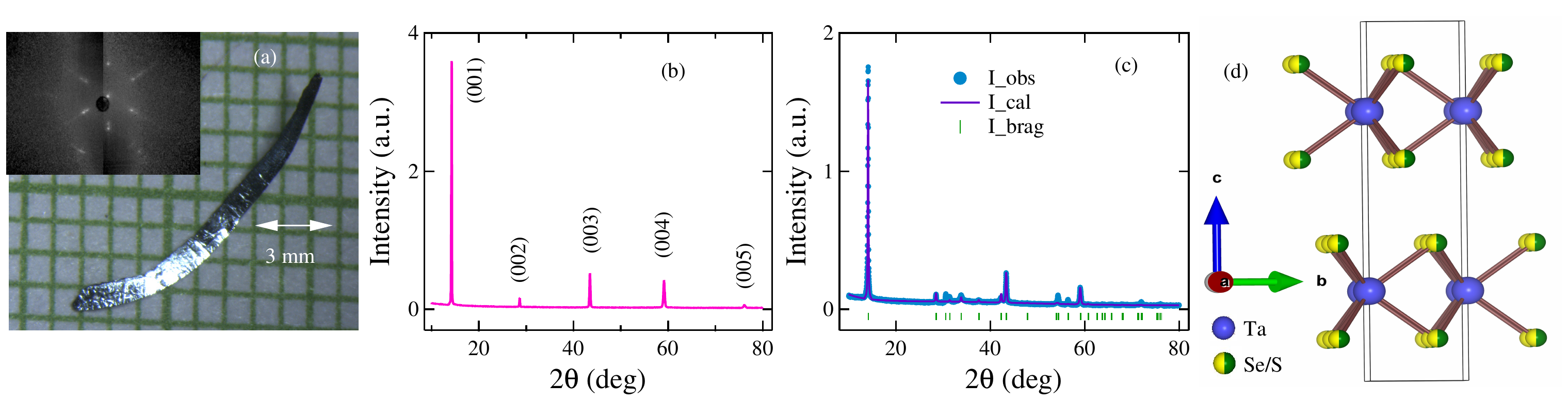}
 \caption{\label{XRD} (a) The microscopic image of single crystal and the inset shows the Laue image of the single crystal. (b) Single crystal powder X-ray diffraction pattern indicates crystal oriented along [00l] plane. (c) Refinement of powder sample for determining lattice parameters. (d) Side view of 2H-TaSeS unit cell.}
\end{figure*}

\section{EXPERIMENTAL METHODS}
2H-TaSeS single crystals were synthesized by the chemical vapour transport method using iodine as the transport agent. The stoichiometric ratios of Ta, Se, and S were thoroughly ground and sealed in a quartz ampule with iodine as the transport agent. The ampule was then placed in a two-zone furnace at a temperature gradient of 850$^{\circ}$C - 750$^{\circ}$C. After 15 days, shiny single crystals were grown in the cold zone of the tube. The Laue diffraction pattern was recorded using a Photonic-Science Laue camera. The orientation of the crystal plane was also confirmed by X-ray diffraction (XRD) at room temperature on a PANalytical diffractometer equipped with $CuK\alpha$ radiation ($\lambda$ = 1.54056 $\text{\AA}$). Magnetization measurements were done using a Quantum Design Magnetic Measurement System (MPMS3, Quantum Design). Transport properties were measured using the four-probe method and specific heat measurements were performed using the two-tau model in the Physical Property Measurement System (PPMS, Quantum Design).

\section{RESULTS AND DISCUSSION}

\subsection{Sample characterization} 
The microscopic image of single-crystal 2H-TaSeS shows crystal in centimetre order in \figref{XRD}(a). The inset of \figref{XRD}(a) represents the Laue pattern of the crystal. It confirms that the crystal orientation is along the [00l] direction. It was further confirmed by the powder XRD \figref{XRD}.(b). To determine the lattice parameters, powder XRD performed on crushed single crystals and confirmed a hexagonal structure with lattice parameters, a = b = 3.37(2) $\text{\AA}$, c = 12.38(4) $\text{\AA}$, which is in agreement with earlier reports of 2H-TaS$_2$ \cite{TaSe2_xSx}. The structure of the unit cell is shown in \figref{XRD}(d).
\begin{figure}
	\includegraphics[width=1.0\columnwidth]{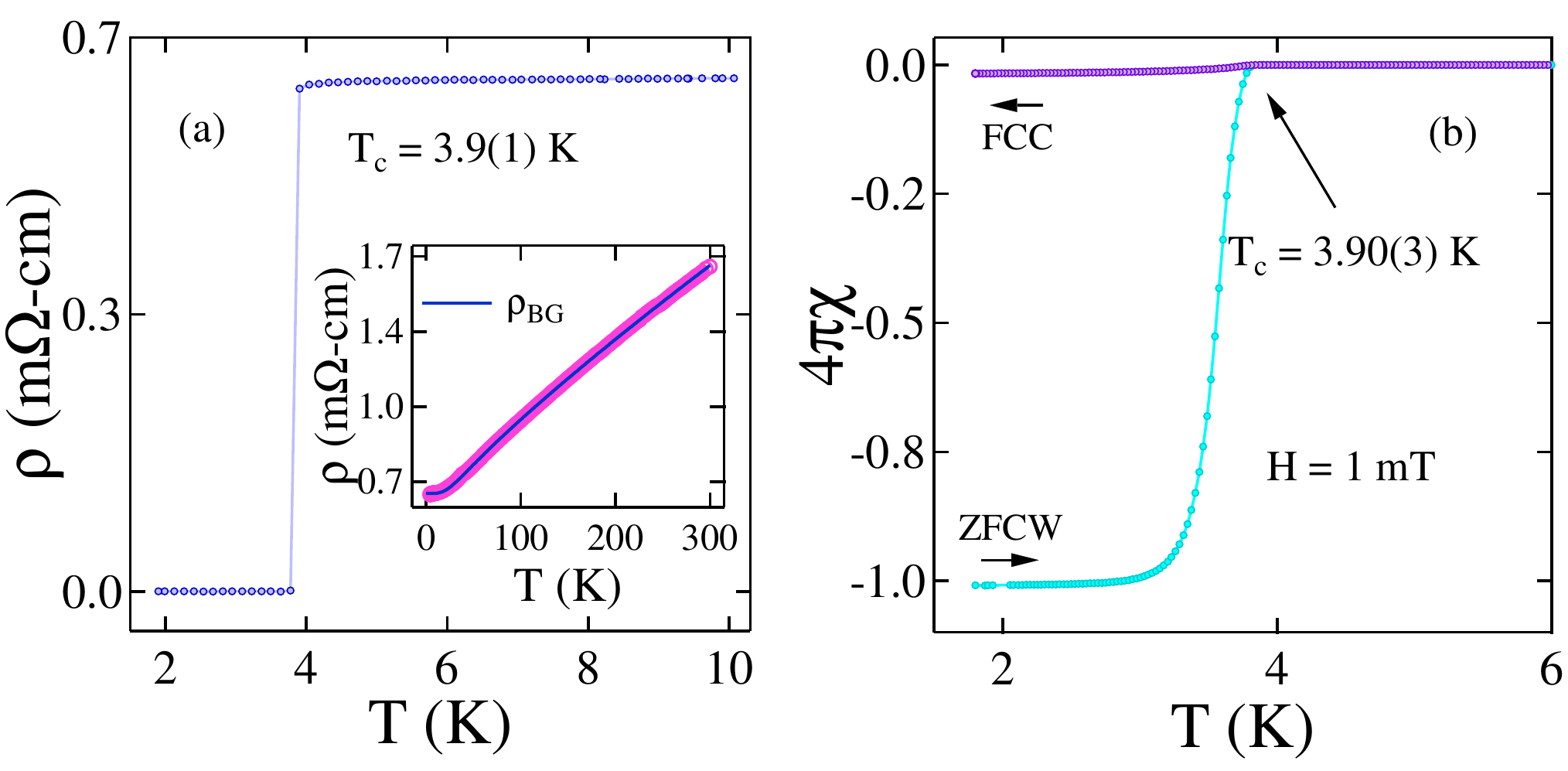}
	\caption{\label{reg and tc} (a) Zero resistivity drop happens at 3.9(1) K and the inset shows the resistivity data above the superconducting transition temperature is fitted by the BG model. (b) Magnetization data was collected in ZFCW-FCC mode that shows the superconducting transition temperature at 3.90(3) K at 1 mT applied field.}
\end{figure}

\subsection{Electrical Resistivity and Magnetization}    
The temperature-dependent resistivity of 2H-TaSeS above 10 K, which shows metallic behaviour up to 300 K and below 3.9(1) K, undergoes a superconducting transition. The resistivity data above the transition temperature was well fitted by using the Bloch-Gr\"{u}neisen model \cite{resistivity_of_MgB2,Bi2Pd,theory_parallel_resistor}. In accordance with this model resistivity, can be described as,
 
 \begin{equation} \label{eq21}
    \rho (T) = \rho_0 + \rho_{BG} (T)
\end{equation}
where $\rho_{BG}(T)$ is defined as
 \begin{equation}  \label{eq2}
\rho_{BG}(T) =  r\left(\frac{T}{\Theta_D}\right)^5 \int_{0}^\frac{\Theta_D}{T} \frac{x^5}{(e^x-1)(1-e^{-x})}dx
\end{equation}
 
 where $\rho_0$ is residual resistivity, $r$ is the material-dependent constant, and $\Theta_D$ is the Debye constant. The fit using Eq.(1) can be seen in inset of \figref{reg and tc}(a) which give $\rho_0$ = 635.42(8) $\mu\ohm$-cm, $r$ = 1.70(2) m$\Omega$-cm, and $\Theta_D$ = 115(1) K.\\
 
\begin{figure}
	\includegraphics[width=1.0\columnwidth]{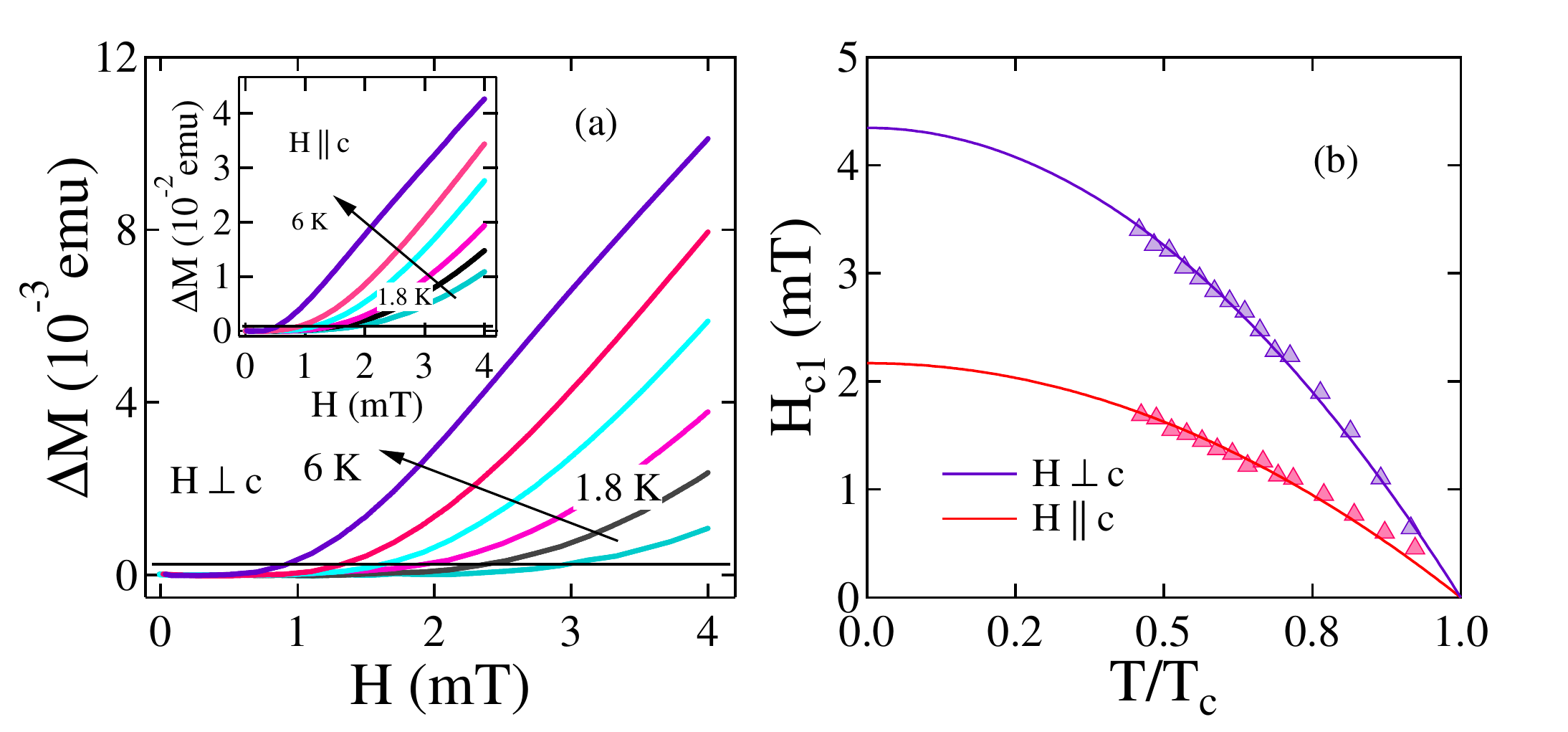}
	\caption{\label{MH&Hc1} (a) Low field variation of magnetization data in two different directions. (b) Lower critical field variation with temperature is well fitted with GL equations and gives values 4.3(2) and 2.1(1) mT for $H \perp c$ and $H \parallel c$ directions.}
\end{figure}

\begin{figure*}[ht!]
	\includegraphics[width=1.8\columnwidth]{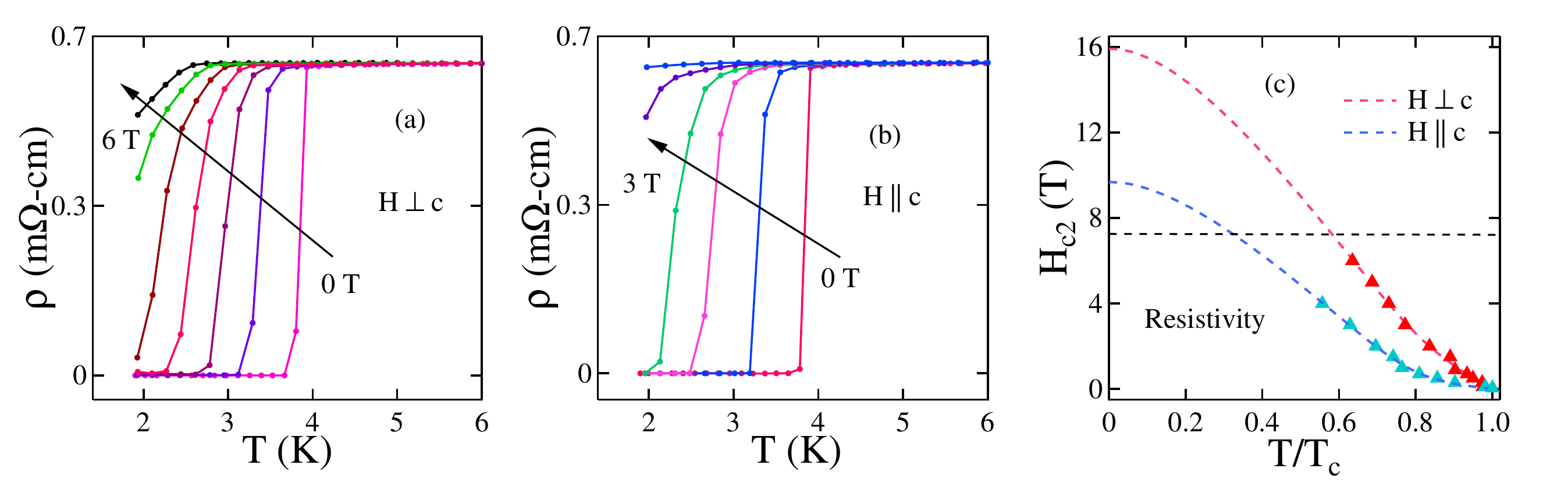}
	\caption{\label{Hc2} (a) and (b) Shows resistivity variation with temperature for $H \perp c$ and $H \parallel c$, respectively. (c) The two-gap fitting of the upper critical field from resistivity data for perpendicular and parallel directions and the black dash lines show Pauli limit of upper critical field, i.e. $H_{c2}^P$ = 1.86$T_c$ =7.25 T.}
\end{figure*}

The superconducting transition of the sample was confirmed by dc magnetization measurement on a single crystal in zero field cooled worming (ZFCW) and field cooled cooling (FCC) mode with an applied 1 mT field. The transition temperature is observed at $T_c$ = 3.90(3) K for $H \perp c$ (\figref{reg and tc}(b)) and $T_c$ = 3.83(3) K for $H \parallel c$. For both perpendicular and parallel field directions, $T_c$ is almost the same. But lower critical field ($H_{c1}(T)$) and the upper critical field ($H_{c2}(T)$) give a huge difference in different directions.  For the calculation of $H_{c1}(0)$ \cite{Two_gap_in_Hc1_in_CaFeCoAsF,Anisotropies_Hc1_MgB2}, the field value for the corresponding temperature is extracted from the magnetization curve (\figref{MH&Hc1}(a)), from which the Meissner lines have deviated. The temperature variation of the lower critical field value is fitted with the Ginzburg-Landau equation by \equref{eqn3}. 
\begin{equation}\label{eqn3}
    H_{c1}(T) = H_{c1}(0)\left[1-\left(\frac{T}{T_c}\right)^2\right]
\end{equation}
\figref{MH&Hc1}(b) clearly shows anisotropy in a different direction, and the value of $H_{c1}(0)$ for $H \perp c$ is 4.3(2) mT and for $H \parallel c$ is 2.1 (1) mT.

The upper critical field ($H_{c2}(T)$) of the 2H-TaSeS system was measured from the transport ($\rho-T$) measurement. The resistivity was measured both in-plane and out-of-plane directions of the crystal. \figref{Hc2}(a) and (b) shows the resistivity with temperature in different directions in magnetic field up to 6.0 T. From the resistivity curves, $H_{c2}(T)$ values were extracted by taking onset value for $T_c$. \figref{Hc2}(c) shows an upturn present near $T_c$ in both the $H \perp c$ and $H \parallel c$ directions. It can not be explained by Ginzburg-Landau theory and Werthamer-Helfand-Hohenberg (WHH) \cite{WHH_Hc2} model. However, this type of behaviour was observed in MgB$_2$ \cite{MgB2_very_high_upper_critical_field,MgAlB2_2_gap,MgBC2_2_gap}, LaFeAsO$_{0.89}$F$_{0.11}$ \cite{Two_Band_Sc_LaFeAsOF}, and some iron-based superconductors, which can be fitted using the two-band model\cite{MgB2_two_gap_theory,Hole_pocket_driven_superconductivity,Twoband_of_112_iron,clean_or_dity}. It can be expressed as

\begin{equation}\label{eqn41}
\begin{split}
ln\frac{T}{T_c} = \frac{1}{2}\left[U(s) + U(\eta s) + \frac{\lambda_0}{w}\right] -\\ \{\frac{1}{4}\left[U(s) -
U(\eta s) - \frac{\lambda_-}{w}\right]^2 + \frac{\lambda_{12}\lambda_{21}}{w}\}^\frac{1}{2}
\end{split}
\end{equation}
$$
 H_{c2} = \frac{2\phi_0T_S}{D_1}; \eta = \frac{D_2}{D_1} $$
 and
$$
 U(s) = \psi(s + \frac{1}{2}) - \psi(\frac{1}{2})
$$
where $\lambda_- = \lambda_{11} - \lambda_{22}$, $\lambda_0 = (\lambda_-^2 + 4\lambda_{12}\lambda_{21})^{1/2}$ and $w = \lambda_{11}\lambda_{22} - \lambda_{12}\lambda_{21}$.  $\lambda_{11}$, $\lambda_{22}$ are the intraband coupling constants and $\lambda_{12}$, $\lambda_{21}$ are the interband coupling constants. $D_1$ and $D_2$ are diffusivities of two bands, respectively. $\phi_0$ is flux quantum and $\psi(x)$ is the digamma function. \figref{Hc2}(c) represents the multi-gap feature with anisotropy in $H \perp c$ and $H \parallel c$ directions. The fit using \equref{eqn41} gives the upper critical field value $H_{c2}^{\perp}(0)$ 15.97(3) T for perpendicular and $H_{c2}^{\parallel}(0)$ 9.59(1) T for parallel directions of field.

In a type-II superconductor, the Cooper pair can break in the applied magnetic field via the orbital and Pauli paramagnetic limiting. In orbital pair breaking, the field-induced kinetic energy of a Cooper pair exceeds the superconducting condensation energy, whereas in Pauli paramagnetic limiting it is energetically favourable for the electron spins to align with the magnetic field, thus breaking the Cooper pairs. For the type-II superconductor, Pauli paramagnetic or Clogston-Chandrasekar limit is defined as $H_{c2}^P$ = 1.86$T_c$. 

\begin{table}[h!]
\caption{Comparative superconducting parameters for 2H-TaSeS with some layered compounds}
\label{hc_parameter}
\begin{center}
\begin{tabular*}{1.0\columnwidth}{l@{\extracolsep{\fill}}cccc}\hline\hline
Parameter& TaSeS & NbSe$_2$ \cite{NbSe2_pauli} & NbS$_2$ \cite{NbS2_pauli} & Ba$_6$Nb$_{11}$Se$_{28}$ \cite{Ba6Nb11Se28}\\
\hline
\\[0.5ex]
$H_{c2}^{\perp}$(T) & 15.97 & 17.3 & &8.84\\
$H_{c2}^{\parallel}$(T) & 9.59 & 5.3 & 1.6 & 0.57\\
$H_{c2}^{P}$(T) & 7.25 & 13.54 & 10.4 & 4.27\\
$\xi_{\perp c}(\text{\AA}$) & 58.61 & 78.8 & 143 & 240.4\\
$\xi_{\parallel c}(\text{\AA}$) & 35.19 & 24.1 & 9.7 & 15.6\\
$\gamma$ & 1.67 & 3.3 & 7.94 & 10.53\\
\\[0.5ex]
\hline\hline
\end{tabular*}
\par\medskip\footnotesize
\end{center}
\end{table}
The Pauli violation ratio defined as $H_{c2}^{^{\parallel}}(0)/H_{c2}^P$. For 2H-TaSeS, PVR in perpendicular and and parallel directions are 2.2 and 1.3 respectively. Violation of Pauli limit of the upper critical field has been observed in other layered superconductors like NbSe$_2$ \cite{NbSe2_pauli}, NbS$_2$ \cite{NbS2_pauli}, and Ba$_6$Nb$_{11}$Se$_{28}$ \cite{Ba6Nb11Se28}, when applied magnetic field is perpendicular to the crystallographic c-axis [see \tableref{hc_parameter}]. However, in case of 2H-TaSeS this violation has been observed for both in-plane ($H \perp c$) and out-of-plane ($H \parallel c$) directions with an anisotropy factor ($\gamma =H_{c2}^{\perp}(0)/H_{c2}^{\parallel}(0)$) of 1.67 \cite{Nb2PdSe5,Ba6Nb11Se28,Fe_based_hc2}. In the layered superconductors Pauli limit voilation can happen due to strong spin-orbit coupling or finite-momentum pairing \cite{bulk_2D_SC,2D_material,graphene}. Strong SOC leads Ising type superconductivity, which is observed recently in monolayer 2H-NbSe$_2$ \cite{MoS2,ising_nbse2}. The finite-momentum pairing can gives rise to the Fulde-Ferrell-Larkin-Ovchinnikov (FFLO) state. To certain the exact mechanism of Pauli limit violation, further low temperature angle depend measurements along with theoretical inputs are required.\\

\begin{figure}
	\includegraphics[width=1.0\columnwidth]{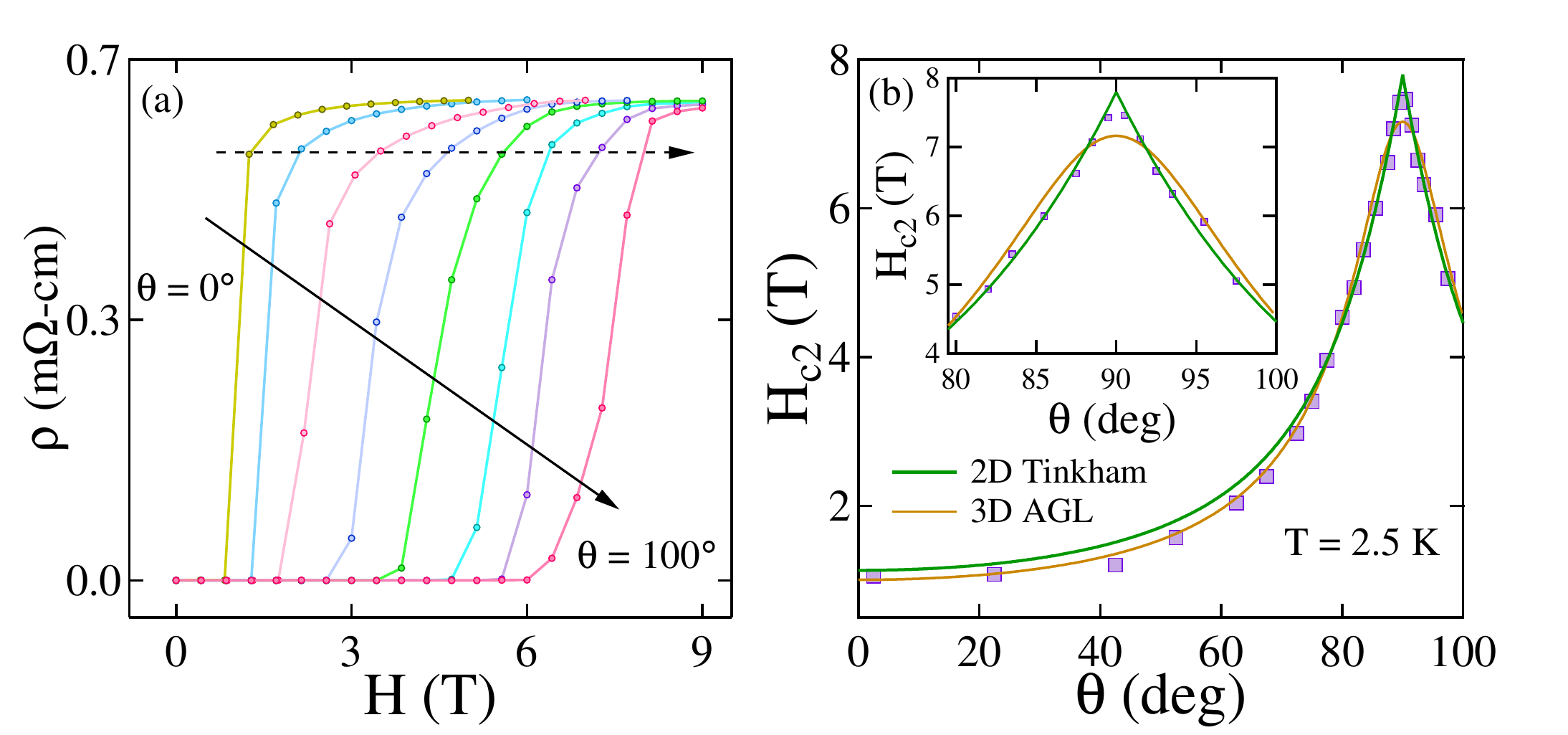}
	\caption{\label{Hc2_angle_dependence} (a) The angle dependence resistivity variation indicates anisotropy in the system. (b) The angle dependence upper critical field from resistivity measurement 2.5 K fitted with 3D GL model and 2D Tinkham model.}
\end{figure}

To further explore the anisotropy, in-plane and out-of-plane upper critical fields, the resistivity was measured at different angles. The angle variation of field dependence of the resistivity at 2.5 K \cite{superconductor_CeIr3} is shown in \figref{Hc2_angle_dependence}(a), where $\theta$ is the angle between the magnetic field and the normal to the sample plane. The angle dependence of resistivity curve shows a clear anisotropy signature from angles $\theta = 0\degree$ ($H \parallel c$) to $\theta = 90\degree$ ($H \perp c$). The upper critical field ($H_{c2}(\theta,T)$) data in \figref{Hc2_angle_dependence}(b) shows a cusplike peak at $\theta \approx 90\degree$. This angular dependence of cusp-like feature can be explained by two models, i.e. 3D anisotropic GL (AGL) model and the model for thin film, 2D Tinkham model \cite{3D_GL_model,Sr2RuO4,AuSn4}. The relevant equations for these models are in \equref{eqn5} and \equref{eqn55}, respectively.
\begin{equation}\label{eqn5}
    \left(\frac{H_{c2}(\theta,T) sin \theta}{H_{c2}^{\perp}}\right)^2 + \left(\frac{H_{c2}( \theta,T) cos\theta}{H_{c2}^{||}}\right)^2 = 1
\end{equation}
\begin{equation}\label{eqn55}
    \left(\frac{H_{c2}(\theta,T) sin \theta}{H_{c2}^{\perp}}\right)^2 + \left\arrowvert\frac{H_{c2}( \theta,T) cos\theta}{H_{c2}^{||}}\right\arrowvert = 1
\end{equation}
Solid orange and green lines show the 3D GL and 2D Tinkham model fitting consequently. The 2D Tinkham model gives a better fitting result compared to others represented in the inset of \figref{Hc2_angle_dependence}(b). This implies that superconductivity is better described as 2D than as 3D for this sample. This 2D kind of superconducting behaviour was observed in 4H$_b$TaS$_2$ \cite{TRSB_4H_TaS2}.\\

The Ginzburg-Landau coherence length ($\xi_{\perp c}$(0) = 58.61(2)  $\text{\AA}$ and $\xi_{\parallel c}$(0) = 35.19(2)  $\text{\AA}$) estimate using $H_{c2}^{\parallel c}(0) = \frac{\phi_0}{2\pi\xi_{\perp c}^2(0)} $ and $ H_{c2}^{\perp c}(0) = \frac{\phi_0}{2\pi\xi_{\parallel c}(0)\xi_{\perp c}(0)}$. Using the Ginzburg-Landau coherence lengths and the lower critical field values ($H_{c1}^{\perp}(0)$ = 4.3(2) mT and $H_{c1}^{\parallel}(0)$ = 2.1(1) mT), the penetration depth of Ginzburg-Landau ($\lambda_{\perp c}(0)$ = 6115(13) \text{\AA} and $\lambda_{\parallel c}(0)$ = 2821(11) \text{\AA}) is calculated using \equref{eqn7}$-$\eqref{eqn71} \cite{LiFeAs_Hc1}.

\begin{equation}\label{eqn7}
    H_{c1}^{\parallel c}(0) = \frac{\phi_0}{4 \pi \lambda^2_{\perp c}(0)}ln\left[\frac{\lambda_{\perp c}(0)}{\xi_{\perp c}(0)} + 0.12\right]
\end{equation}

\begin{equation}\label{eqn71}
    H_{c1}^{\perp c}(0) = \frac{\phi_0}{4 \pi \lambda_{\parallel c}\lambda_{\perp c}(0)}ln\left[\frac{\lambda_{\parallel c}(0)}{\xi_{\parallel c}(0)} + 0.12\right]
\end{equation}

where $\phi_0$ (= 2.07$\times$10$^{-15}$ T$m^2$) is the magnetic flux quantum. The two characteristic length parameters are used to calculate the Ginzburg-Landau parameter $\kappa_{\perp c}$ = 104 and $\kappa_{\parallel c}$ = 80$>$ $\frac{1}{\sqrt{2}}$ by \equref{eqn8}, indicating a type II behaviour of the sample.
 \begin{equation}\label{eqn8}
     \kappa_{\perp c} = \frac{\lambda_{\perp c}(0)}{\xi_{\perp c}(0)}
 \end{equation}

\subsection{Specific heat}    

The zero-field low temperature specific heat data show a discontinuity at 3.79(7) K, the same as the reported superconducting transition temperature via resistivity and magnetization measurements. The zero-field data above the superconducting transition temperature is fitted with $C/T = \gamma_n + \beta_3 T^2$, where $\gamma_n$ is the electronic contribution, and $\beta_3$ is the phononic contribution. The parameters were 8.13 mJ/mol-K$^2$ and 1.33 mJ/mol-K$^4$, respectively. Furthermore, the Debye temperature, $\theta_D$ = 163.34 K, was calculated using \equref{eqn9}.
\begin{equation} \label{eqn9}
    \theta_D = \left(\frac{12\pi^4RN}{5\beta_3}\right)^{\frac{1}{3}}
\end{equation}
where $R$ is the universal gas constant (=8.314 J mol$^{-1} K^{-1}$) and $N$ is the number of atoms per formula unit.\\

To reveal the superconducting gap parameter, the detailed electronic specific heat ($C_{el}(T)$) in the superconducting state analysed. The $C_{el}(T)$ (\equref{eq12}) in the superconducting state was calculated by subtracting the phononic contribution from total specific heat $C(T)$. 
\begin{equation}\label{eq12}
    C_{el}(T) = C(T) - \beta_3T^3
\end{equation}

\begin{figure}
	\includegraphics[width=1.0\columnwidth]{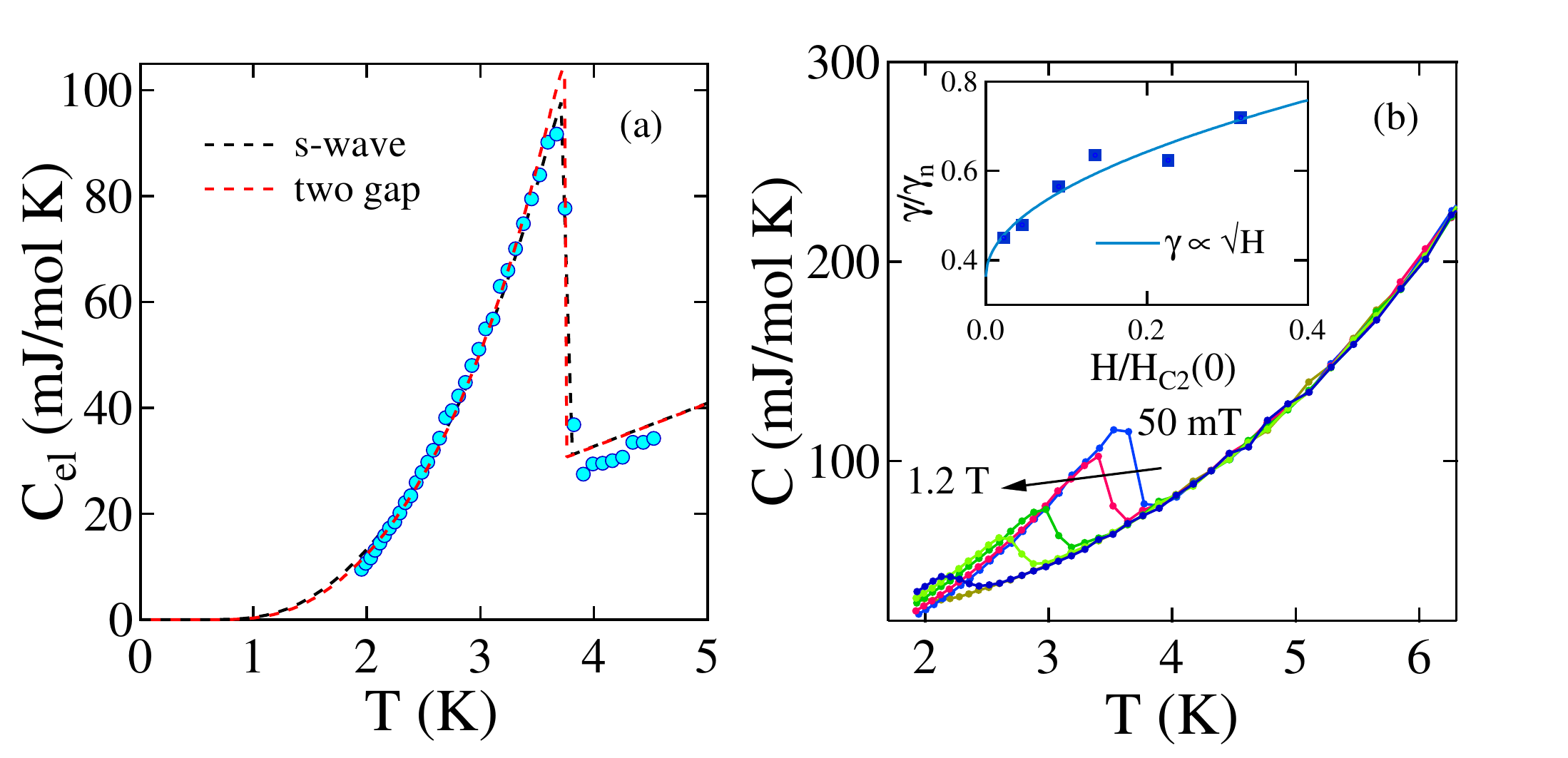}
	\caption{\label{Cel_fitting_with layout.jpg} (a) low-temperature electronic contribution of specific data is better fitted with two-gap comparative single gap model. (b) The Sommerfeld coefficient was calculated from field-dependent of specific heat curve. The inset shows $\gamma$ depends on the square root of H.}
\end{figure}

The temperature dependence of the electronic contribution of specific heat in the superconducting state can be described using the fully gaped model given below in \equref{eq13}.

\begin{equation}\label{eq13}
    \frac{S}{\gamma_nT_c} = -\frac{6}{\pi^2}\left(\frac{\Delta(0)}{k_BT_c}\right)\int_{0}^{\infty} [fln(f) + (1-f)ln(1-f)] \,dy,
\end{equation}
where $f(\xi) = [exp(E(\xi)/k_BT) + 1]^{-1}$ is the fermi function $E(\xi) = \sqrt{\xi^2 + \Delta(t)^2}$, where $\xi$ is energy of normal electron comparative to Fermi energy, $y =\xi/\Delta(0)$ , $t = T/T_c$ and $\Delta(t) = tanh[1.82(1.018((1/t)-1))^{0.51}]$ is the
BCS approximation for the temperature dependence of the
energy gap. The normalized electronic specific heat is related
to entropy by \equref{eq14}.

\begin{equation} \label{eq14}
    \frac{C_{el}}{\gamma_nT_c} = t\frac{d(S/{\gamma_n T_c})}{dt}
\end{equation}

The temperature dependence of $C_{el}(T)$ was fitted using \equref{eq13} and \equref{eq14}. In \figref{Cel_fitting_with layout.jpg}(a), the dotted black line represents the s-wave fitting and gives the gap value of 2.19 meV. It reproduces the experimental data above T $\approx$ 2.38 K. The s-wave fitting deviates at lower temperatures. To explain the temperature behaviour, the phenomenological two-gap $\alpha$ model (s+s wave) is used \cite{Strong_Coupled_Superconductors,theory_two_gap}. In this model, each band is characterized by the corresponding Sommerfeld coefficient $\gamma_n$ = $\gamma_1$ + $\gamma_2$, and the total specific heat was calculated by two gap parameters ($\Delta_1$ and $\Delta_2$) and their comparative weights ($\gamma_1/\gamma_n \equiv$ x and $\gamma_2/\gamma_n \equiv$ 1-x). The \figref{Cel_fitting_with layout.jpg}(a) (red dotted line) exhibits a better agreement across the whole temperature range, in particular for T < 2.38 K. The gap values are 1.49 meV and 2.27 meV, with a fraction of 0.09. To obtain the true nature of the superconducting gap, heat-capacity data is to be analyzed well below $T_c$/10.

The superconducting gap symmetry can be further confirmed by the magnetic field dependence of the Sommerfeld coefficient $\gamma$(H). In conventional fully gapped type-II superconductor it is  proportional to the vortex density. As we apply more
field the vortex density increases due to an increase in the
number of field induced vortices which in turn enhances the
quasiparticle density of states. This gives rise to a linear
relation between $\gamma$ and H, i.e., $\gamma(H)$ $\propto H$ for a nodeless and isotropic s-wave superconductor \cite{Volovik_effect,Behavior_of_dirty_superconductors,Superconductivity_of_Metals_and_Alloys}. For a superconductor with nodes in the gap, Volovik predicted a nonlinear relation
given by $\gamma(H)$ $\propto$ $\sqrt{H}$ \cite{d_Wave_Superconductors,YNi2B2C,N_Nakai_et_al,specific_heat_2_gap_NbSe2}. The Sommerfeld coefficient $\gamma$ was calculated by fitting with \equref{eq15} for various fields and
extrapolating it to T = 0 K [see \figref{Cel_fitting_with layout.jpg}(b)]. 
\begin{equation}\label{eq15}
\frac{C_{el}}{T} = \gamma + \frac{a}{T}exp\left(-b\frac{T_c}{T}\right)
\end{equation}
The inset \figref{Cel_fitting_with layout.jpg}(b) shows that the Sommerfeld coefficient varies with the square root of H. The linear deviation of $\gamma$ indicates that TaSeS is a possible multi-gap system.

\begin{figure}
	\includegraphics[width=1.0\columnwidth]{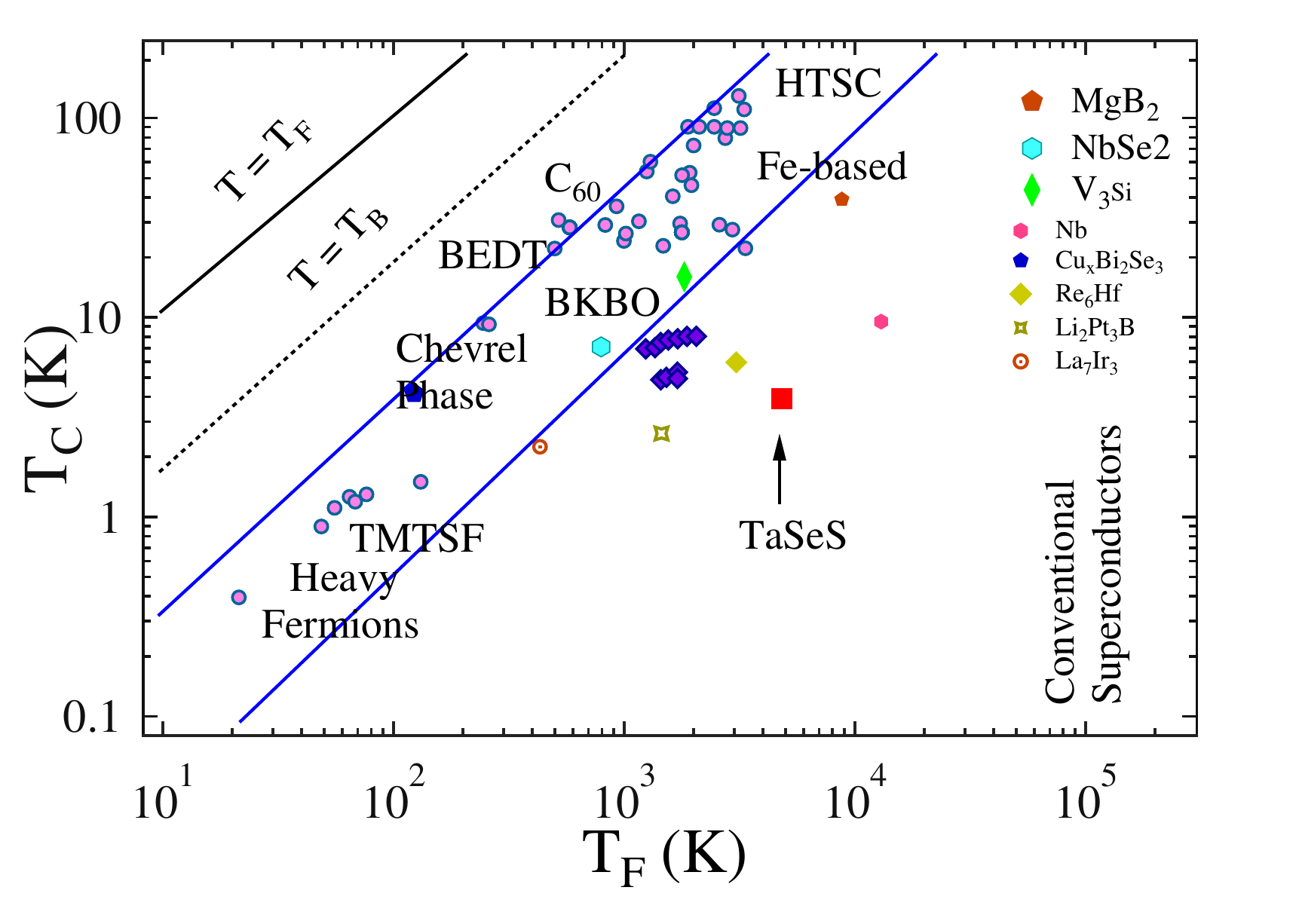}
	\caption{\label{uemra} The plot of the superconducting transition temperature versus Fermi temperature for different superconducting families. In between, two solid blue lines show the unconventional band of superconductors. TaSeS lies close to the unconventional band.}
\end{figure}

The superconducting materials can be classified as conventional or unconventional based on T$_c$/T$_F$ ratio that is given by Uemura et al. \cite{1st_uemra,2nd_uemra,3rd_uemra}. The Chevrel phases, heavy fermions, Fe-based superconductors, and high T$_c$ superconductors are fallen into the unconventional category as T$_c$/T$_F$ ratio within 0.01$\le$T$_c$/T${_F}\le$0.1 range.  Fermi temperature for TaSeS is  4825 K, obtained by solving a set of five equations \cite{uemra_equation,uemra_equation_2,uemra_plot_2} simultaneously. The  ratio of T$_c$/T$_F$ = 0.0008 place it  2H TaSeS near to the band of unconventional superconducting materials [see \figref{uemra}].

\section{CONCLUSION}

In summary, we studied the magnetization, electrical, and magnetotransport properties of 2H-TaSeS. It confirms multigap superconductivity in 2H-TaSeS  having a superconducting transition temperature of 3.9 K and Pauli limit breaking in both the in-plane and out-of-plane directions. The angle dependence of the upper critical field well-fit 2D Tinkham model suggests 2D superconductivity in bulk 2H-TaSeS single crystals. All results indicate 2H-TaSeS a is a new candidate for unconventional superconductor and can host Ising or FFLO type superconductivity. Further, low temperature and thickness dependence measurements along with theoretical inputs are required to understand the exact superconducting pairing mechanism. 

\section{Acknowledgments}

R.~P.~S.\ acknowledge Science and Engineering Research Board, Government of India, for the CRG/2019/001028 Core Research Grant.

\end{document}